\documentclass[twocolumn,aps,prl,superscriptaddress]{revtex4}
\usepackage{amssymb}
\usepackage{amsmath,bm}
\usepackage{graphicx}
\usepackage[normalem]{ulem}
\usepackage{multirow}
\usepackage[colorlinks,linkcolor=blue,urlcolor=blue,citecolor=blue]{hyperref}
\usepackage{lipsum}
\usepackage[usenames, dvipsnames]{xcolor}
\usepackage{tensor}
\usepackage{isotope}
\usepackage{amsmath}
\allowdisplaybreaks[4]
 
\renewcommand{\sout}{\bgroup \color{red} \ULdepth=-.5ex \ULset}

\begin{document}

\title{Event-by-event anti-deuteron multiplicity fluctuation in Pb+Pb collisions at  $\sqrt{s_{NN}}=5.02~$TeV}

\author{Kai-Jia Sun}
\email{kjsun@tamu.edu}
\affiliation{Cyclotron Institute and Department of Physics and Astronomy, Texas A\&M University, College Station, Texas 77843, USA}

\author{Che Ming Ko}
\email{ko@comp.tamu.edu}
\affiliation{Cyclotron Institute and Department of Physics and Astronomy, Texas A\&M University, College Station, Texas 77843, USA}
 
\date{\today}
\begin{abstract}
Using the nucleon coalescence model based on kinetic freeze-out nucleons from the hybrid model of MUSIC hydrodynamics and UrQMD hadronic transport, we study the production of anti-deuteron and its event-by-event fluctuation in Pb+Pb collisions at $\sqrt{s_{NN}}=5.02$ TeV. We find a clear suppression of the  anti-deuteron to antiproton yield ratio in peripheral collisions, which is in accordance with the measurements from the ALICE Collaboration.   Also found is a Poissonian event-by-event fluctuation of the anti-deuteron multiplicity distribution in all collision centralities, which is different  from the prediction of a simple coalescence model calculation  that assumes all antiproton and antineutron pairs to have the same probability to form anti-deuterons.  We further find a small negative correlation between the anti-deuteron and antiproton multiplicity distributions  as a result of the baryon conservation.
\end{abstract}

\maketitle
\emph{Introduction.}{\bf ---}
Fragile anti-nuclei like anti-deuteron ($\bar{d}$), anti-helium ($\overline{^3\text{He}}$ and $\overline{^4\text{He}}$), and anti-hypertriton ($\overline{^3_{\bar{\Lambda}}\text{H}}$) have been observed in high-energy nuclear collisions at both RHIC and the LHC~\cite{STARSc328,STARNt473,ALICE:2015rey,STARNtP16}.   In particular, their suppressed production  in collisions of small systems at the LHC, which    have been  attributed to their composite internal structures~\cite{Sun:2018mqq} or the canonical effects~\cite{Vovchenko:2018fiy}, has recently attracted a lot of theoretical and experimental interest~\cite{ALICE:2017xrp,ALICE:2019fee,ALICE:2020foi,ALICE:2021ovi,ALICE:2021mfm,ALICE:2021sdc}.  These loosely bound states  also  have important applications in  the search  for the signals of phase transitions  in the strongly interacting matter created in relativistic heavy-ion collisions~\cite{Sun:2017xrx,Sun:2018jhg,Sun:2020zxy,Shuryak:2018lgd,Shuryak:2019ikv,Zhang:2020ewj}  and  in the dark matter detection in space~\cite{Blum:2017qnn,vonDoetinchem:2020vbj}. Understanding their production mechanisms  in high-energy nuclear reactions  is thus of great importance~\cite{Chen:2018tnh,Braun-Munzinger:2018hat}.

Various theoretical models based on different assumptions have been used   to describe   anti-nuclei  production  in nuclear reactions. While light (anti-)nuclei  are assumed to be thermally produced from hadronization of produced quark-gluon plasma (QGP) in the statistical hadronization model (SHM)~\cite{AndNt561}, they are formed from the recombination of nucleons at kinetic freeze-out in the coalescence model~\cite{SchPRC59,Sun:2018mqq,Bellini:2020cbj}.   Also, the kinetic or transport approach, which takes into consideration of the disintegration and regeneration of (anti-)nuclei during the evolution of the hadronic matter before their decoupling, has  been used to describe the production of (anti-)deuterons and other light (anti-)nuclei in relativistic heavy ion collisions~\cite{DanNPA533,OhPRC76,OhPRC80,OliPRC99,Sun:2021dlz}. These different models have all been found to give almost the same  deuteron and anti-deuteron yields in central Pb+Pb collisions at the LHC~\cite{AndNt561,OliPRC99,Zhao:2020irc,Sun:2021dlz}.  It thus remains unclear if these fragile clusters are directly formed at the chemical freeze-out like in the statistical model or if they are formed  at the  later stage of the hadronic evolution like in the coalescence model.

Apart from the event-averaged  multiplicities of light nuclei, their event-by-event fluctuations also provide a  sensitive probe to their production mechanisms.  Since (anti-)nucleons and (anti-)deuterons in SHM are produced independently from the same emission hypersurface, their number fluctuations are close to the Poisson limit. This is, however, different in the coalescence model, in which  anti-deuterons are formed from the  recombination of antiprotons and antineutrons at the kinetic freeze-out where the system is out of chemical equilibrium. In a simple coalescence model
calculation~\cite{Feckova:2016kjx}, which assumes  that all antiproton and antineutron pairs have the same probability to form  deuterons, deviations from the Poisson limit have been observed in the scaled second moment  of deuteron multiplicity distribution as well as in the correlation between the antiproton and anti-deuteron multiplicity distributions, suggesting that the measurement of event-by-event fluctuation of  anti-deuteron multiplicities can provide a sensitive probe to  its production mechanism  in relativistic heavy ion collisions. 

In the present study, we adopt the hybrid model~\cite{Zhao:2021dka} of MUSIC+URQMD+COAL to  study quantitatively the production of anti-deuteron, the event-by-event fluctuation of anti-deuteron multiplicity, and the correlation between the antiproton and antideuteron multiplicities in Pb+Pb collisions at $\sqrt{s_{NN}}=5.02$~TeV.  After the evolution of the QGP produced in these collisions via the use of the (3+1)-dimensional viscous hydrodynamic model  MUSIC~\cite{PaqPRC93,SCPRC97,SCNST31}, hadrons are produced  from a constant energy density particlization hypersurface according to the Cooper-Frye formula~\cite{CooPRD10}. The subsequent hadronic rescatterings and decays are modelled by the URQMD. The anti-deuteron yield in a collision event is then evaluated from the coalescence (COAL) of antiprotons and antineutrons at the kinetic freeze-out.  Volume fluctuation and thermal smearing, which are neglected in the simple coalescence  calculation of Ref.~\cite{Feckova:2016kjx}, are  automatically included in our study.  Also, the baryon conservation is imposed in our coalescence calculation.
 
\begin{figure}[!t]
  \centering
 \includegraphics[width=8cm]{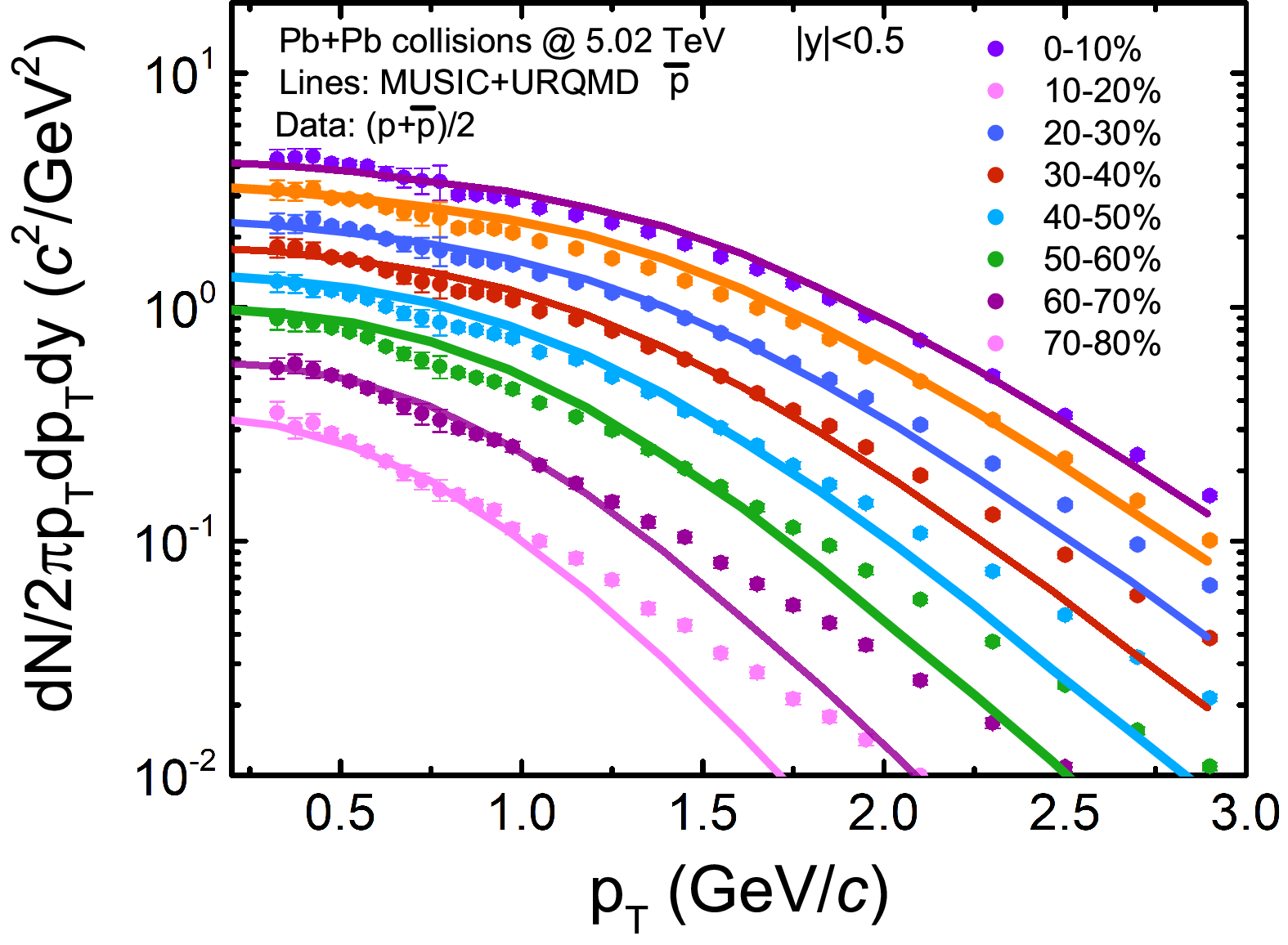}
  \caption{Transverse momentum spectra of antiprotons in Pb+Pb collisions at $\sqrt{s_{\rm NN}}=5.02~\rm TeV$  from theoretical predictions based on MUSIC+URQMD (solid lines) and the experimental data (solid circles)~\cite{ALICE:2019hno}. }
  \label{pic:PT}
\end{figure}

\emph{Suppression of anti-deuteron production in peripheral collisions.}{\bf ---}We first show in Fig.~\ref{pic:PT} by solid lines the transverse momentum spectra of anti-protons obtained from MUSIC+URQMD from collisions at various centralities. Compared to the experimental data from Ref.~\cite{ALICE:2019hno}, shown by  solid circles, the model is seen to describe very well the data in the momentum range of $0.4<p_T<0.9$ GeV/$c$, which covers the kinematic region needed in the present study. With the phase-space  distributions  of kinetic freeze-out antiprotons and antineutrons from $2\times 10^6$ events, we then calculate the anti-deuteron yield from each collision event  using a realistic coalescence model. In this model~\cite{SchPRC59,Sun:2018mqq}, the formation probability of anti-deuteron in the kinetically freeze-out hadronic matter is given by the Wigner function of its internal wave function, which we take as
\begin{eqnarray}
W_{\bar{d}}&=&8g_{\bar{d}}\exp\left({-\frac{x^2}{\sigma_{\bar{d}}^2}}-\sigma_{\bar{d}}^2 p^2\right),   \label{Eq:wig}
\end{eqnarray}
with $g_{\bar{d}}=3/4$ being the statistical factor for spin 1/2 proton and neutron to from a spin 1 deuteron, and the relative coordinate and momentum defined as ${\bf x}=({\bf x}_1-{\bf x}_2)/\sqrt{2}$, ${\bf p}=({\bf p}_1-{\bf p}_2)/\sqrt{2}$. The size parameter in the Wigner function is related to the (anti-)deuteron root-mean-squared radius by $\sigma_{\bar{d}}  = \sqrt{4/3}~r_{\bar{d}}\approx 2.26$ fm~\cite{Sun:2017ooe,Ropke:2008qk}.

\begin{figure}[!t]
  \centering
 \includegraphics[width=8.5cm]{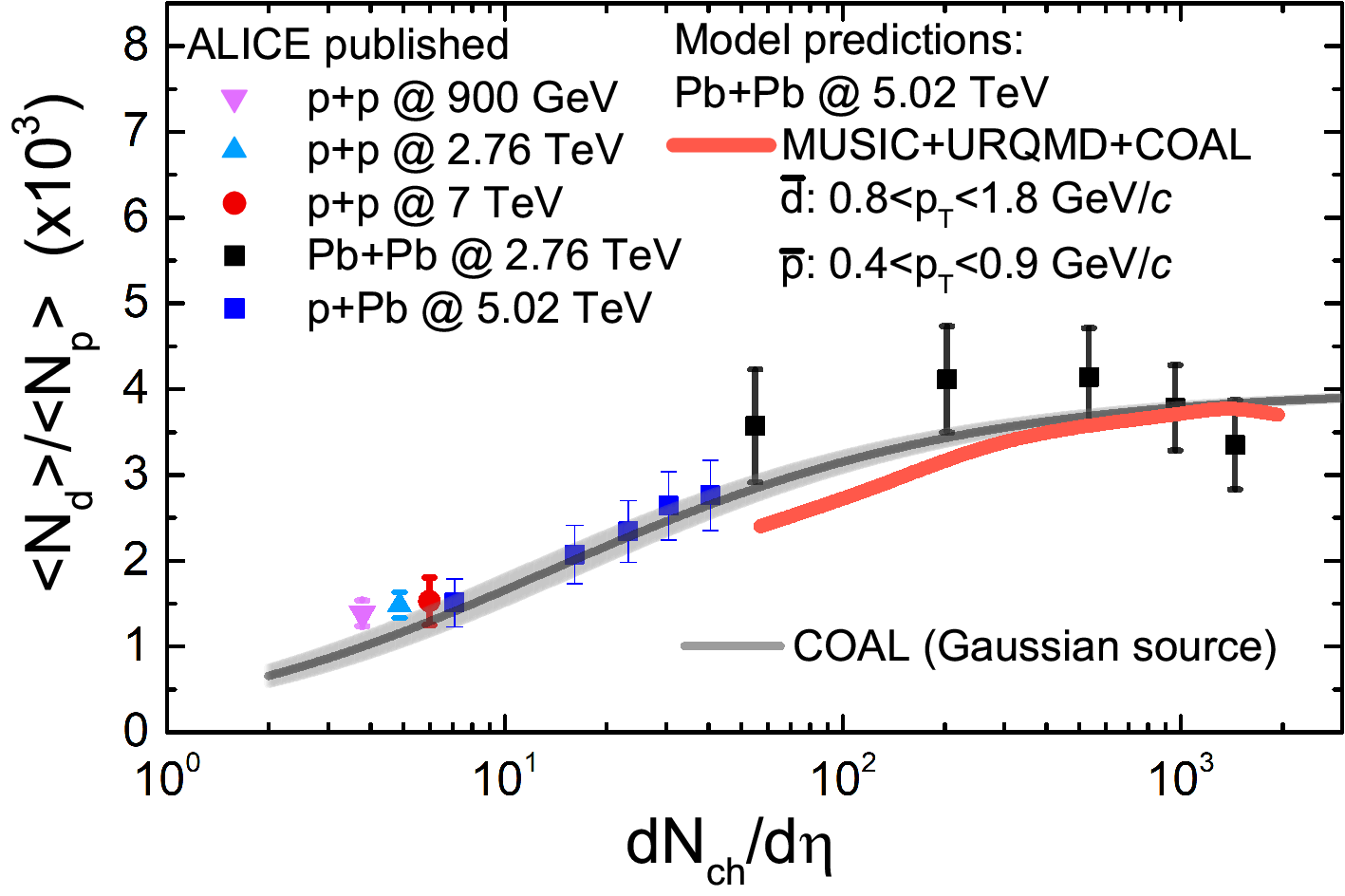}
  \caption{Charged particle multiplicity dependence of the yield ratio $\langle N_{ d}\rangle/\langle N_{ p}\rangle$ or $\langle N_{\bar d}\rangle/\langle N_{\bar p}\rangle$.  The red solid line denotes the prediction of MUSIC+URQMD+COAL. The line with shaded band denotes the prediction of coalescence model with a parametrized Gaussian source~\cite{Sun:2018mqq}. Experimental data from the ALICE Collaboration are shown by symbols with error bars~\cite{ALICE:2015wav,ALICE:2017xrp,ALICE:2019fee,ALICE:2020foi,ALICE:2021ovi,ALICE:2021mfm,ALICE:2021sdc}.}
  \label{pic:d2p}
\end{figure} 

Since the conservation of baryon number is important for studying the event-by-event light-nuclei multiplicity fluctuations, the usual perturbative calculation in the coalescence model, which allows a nucleon to be both a nucleon and also the constituent of light nuclei, needs to be modified, and this is done as follows. For an event with $N_{\bar{p}}$ anti-protons and $N_{\bar{n}}$ anti-neutrons, we first evaluate the coalescence probabilities for all $N_{\bar{p}}\times N_{\bar{n}}$ possible antiproton and antineutron pairs. For each pair, we then determine whether an anti-deuteron is produced according to its formation probability. Once an anti-deuteron is produced, the two anti-nucleons are no longer allowed to form other anti-deuterons  by setting their probability to form another anti-deuteron to zero. In this case, anti-deuterons are  produced sequentially, and the baryon number is exactly conserved  during the production of anti-deuterons.   This method of implementing baryon number conservation is different from that adopted in Ref.~\cite{Feckova:2016kjx}, which assumes  antiproton and antineutron pairs have the same probability to form anti-deuterons, and  they are then subtracted  from the final antinucleon numbers. Although both methods lead to a negative correlation between  the antiproton and anti-deuteron distributions, there is a visible difference in their predicted event-by-event fluctuation in the anti-deuteron multiplicity. 
 
Due to the vanishing baryon chemical potential,  deuterons and anti-deuterons are equally produced in Pb+Pb collisions at $\sqrt{s_{NN}}=5.02$ TeV considered in the present study. Fig.~\ref{pic:d2p} shows the charged particle multiplicity dependence of the yield ratio $\langle N_{ d}\rangle/\langle N_{ p}\rangle$ or equivalently $\langle N_{\bar d}\rangle/\langle N_{\bar p}\rangle$ with  $\langle\dots\rangle$ denoting the average over events.  The line with shaded band denotes the prediction of coalescence model with a parametrized Gaussian emission source~\cite{Sun:2018mqq}. Experimental data from the ALICE Collaboration are shown by symbols with error bars~\cite{ALICE:2015wav,ALICE:2017xrp,ALICE:2019fee,ALICE:2020foi,ALICE:2021ovi,ALICE:2021mfm,ALICE:2021sdc}. The prediction from the present model calculation is shown by the solid red line. It is seen that the suppression of the  (anti-)deuteron to (anti-)proton yield ratio in collisions of system with small charged particle multiplicities is reproduced by our coalescence model that takes into account explicitly the finite (anti-)deuteron size compared to the source size of (anti)protons and (anti)neutrons.  The suppressed production of (anti-)deuterons relative to that of (anti)protons in small collision systems can also be described by the statistical hadronization model via the introduction of the canonical effect~\cite{Vovchenko:2018fiy} by varying the volume over which the baryon number is conserved.

\begin{figure}[!t]
  \centering
 \includegraphics[width=8cm]{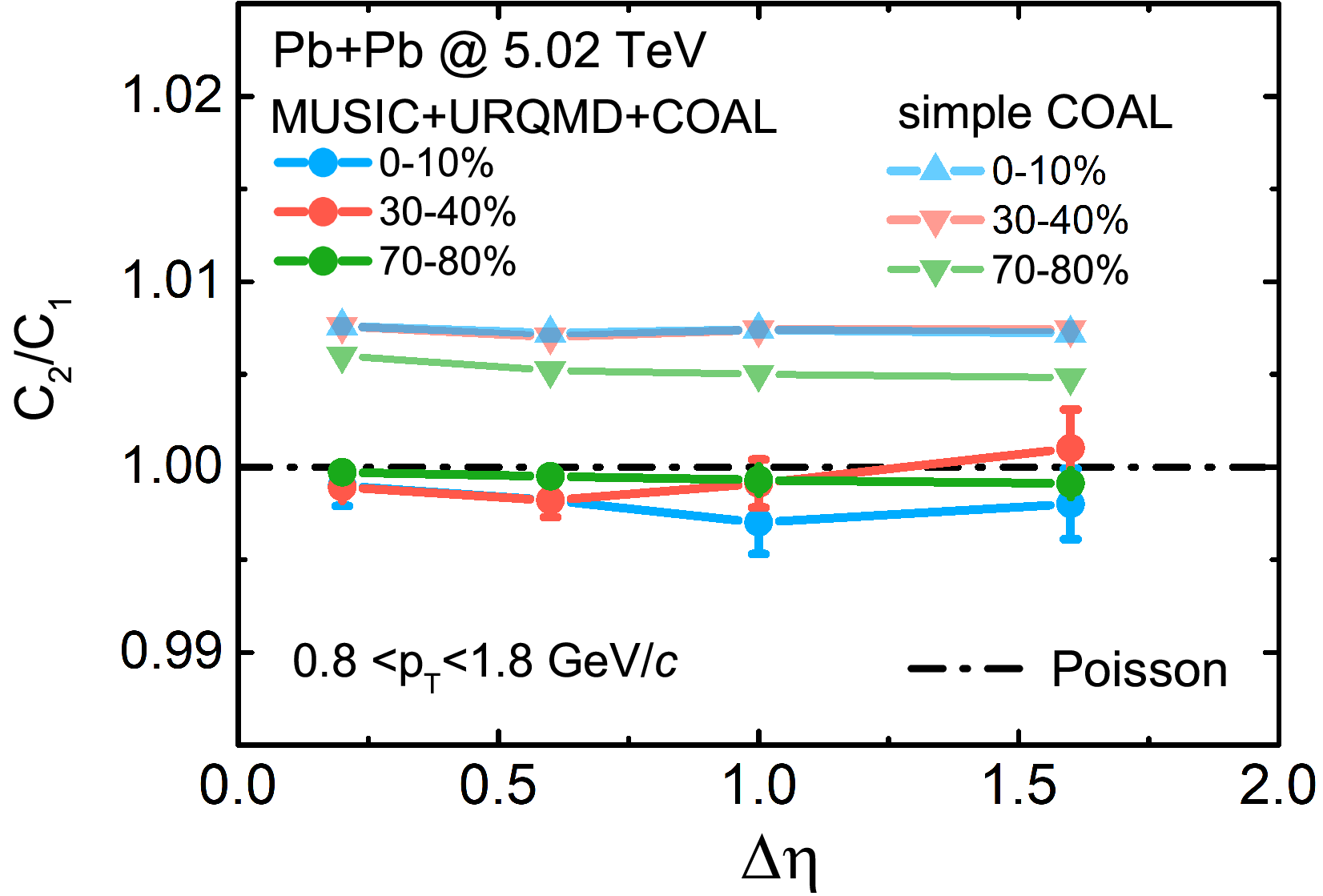}
  \caption{ Scaled moments $C_2/C_1$ of anti-deuteron event-by-event multiplicity distribution as a function of psudo-rapidity acceptance $\Delta \eta$  in Pb+Pb collisions at $\sqrt{s_{NN}}=5.02$~TeV from present study using a realistic coalescence model and from Model B of the simple coalescence model  in Ref.~\cite{Feckova:2016kjx}. The dashed line denotes the Poisson limit.}
  \label{pic:moment}
\end{figure}

\emph{Event-by-event anti-deuteron multiplicity fluctuation}{\bf ---}With the event-by-event multiplicity distributions of antiprotons and anti-deuterons from our model calculation, we now evaluate the ratio of the second moment $C_2=\langle N_{\bar d}^2-\langle N_{\bar d}\rangle^2\rangle$ of the anti-deuteron distribution to its first moment $C_1=\langle N_{\bar d}\rangle$, where $N_{\bar d}$ is the anti-deuteron number or multiplicity in a single event. The results are shown by solid circles in Fig.~\ref{pic:moment}, and they are consistent with the Poisson limit, i.e., $C_2/C_1= 1$,  denoted by the dash-dotted line.   These results are obtained by dividing each centrality bin into 20 smaller bins with equal number of events and weighting the cumulants and correlations calculated in each smaller bin by  the associated average charged particle multiplicity.  This is in contrast to the method of centrality bin width correction (CBWC), which weights the cumulants in each multiplicity bin by the number of events in the bin~\cite{Luo:2013bmi,Chatterjee:2019fey,He:2018mri}.  Also shown by solid triangles are the results from Model B of the simple coalescence model in Ref.~\cite{Feckova:2016kjx}, which gives $C_2/C_1\approx 1+2\langle N_{\bar d}\rangle/\langle N_{\bar p}\rangle$, as shown in the Appendix, and exceeds the Poisson limit.  

We note that the Poisson limit is only reached  if the grand canonical ensemble is used in the statistical hadronization model. Using the canonical ensemble in this model  would reduce the event-by-event multiplicity fluctuation as a result of the baryon conservation~\cite{Vovchenko:2018fiy}. It remains to   be seen how this canonical effect would affect the value of $C_2/C_1$ and the correlation between the antiproton and anti-deuteron multiplicities discussed below.

\begin{figure}[!t]
  \centering
 \includegraphics[width=8cm]{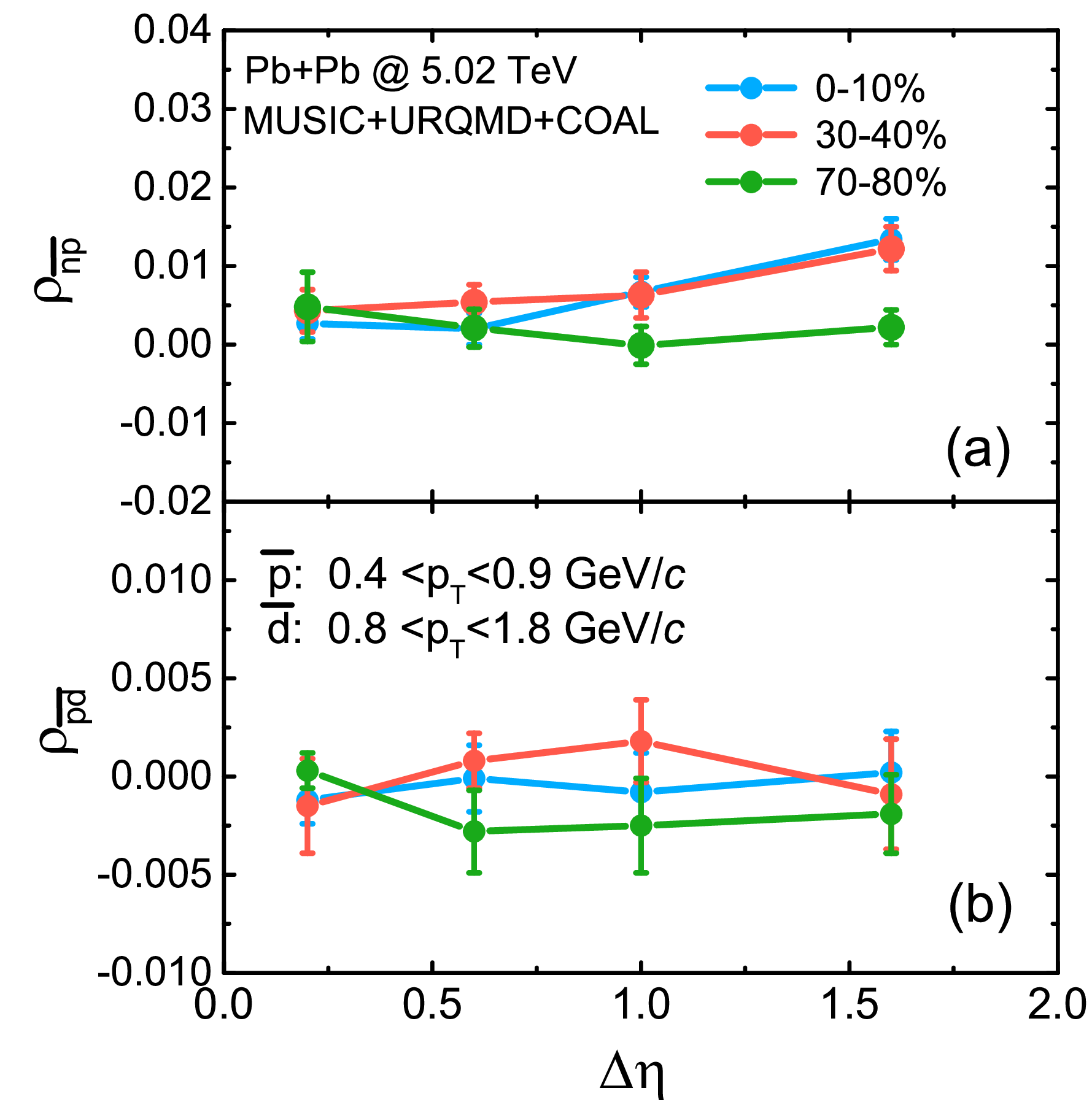}
  \caption{Event-by-event correlation between antineutron and antiproton   multiplicity distributions (panel (a)) and between antiproton and anti-deuteron  multiplicity distributions (panel (b)) as a function of the pseudorapidity acceptance window $\Delta \eta$ in Pb+Pb collisions at $\sqrt{s_{NN}}=5.02$~TeV for different collision centralities.}
  \label{pic:Cnp}
\end{figure}

\emph{Correlation between antiproton and anti-deuteron}{\bf ---}Since an anti-deuteron in the coalescence model is formed  from an antiproton and an antineutron, the correlation between the antiproton and anti-deuteron multiplicity distributions is affected by the correlation between antineutron and antiproton multiplicity distributions.  To explore to what degree the antiproton and antineutron multiplicities in an event are correlated before the coalescence, we evaluate the correlation coefficient $\rho_{\bar{n}\bar{p}}$, defined as
\begin{equation}
\label{corr}
\rho_{\bar{n}\bar{p}} = \frac{\langle(N_{\bar{n}}-\langle N_{\bar{n}}\rangle)(N_{\bar{p}}-\langle N_{\bar{p}}\rangle)\rangle}{\sqrt{\langle N_{\bar{n}}^2-\langle N_{\bar{n}}\rangle^2\rangle\langle N_{\bar{p}}^2-\langle N_{\bar{p}}\rangle^2\rangle}}.
\end{equation}
Panel (a) of Fig.~\ref{pic:Cnp} shows results from our model calculations on the correlation between the antiproton and antineutron multiplicites at the kinetic freeze-out as a function of the pseudorapidity acceptance window $\Delta \eta$ in Pb+Pb collisions at $\sqrt{s_{NN}}=5.02$~TeV for different collision centralities. It is seen that the correlation is close to zero except for central collisions at large pseudorapidity acceptance windows, where the value of $\rho_{\bar{n}\bar{p}}$ is about 0.012.  For the correlation between the antiproton and anti-deuteron multiplicity distributions,  shown in panel (b) of Fig.~\ref{pic:Cnp}, its value is also consistent with zero within uncertainties, which becomes, however, slightly negative for  peripheral collisions at centrality $70-80\%$. 

\begin{figure}[!t]
  \centering
 \includegraphics[width=8cm]{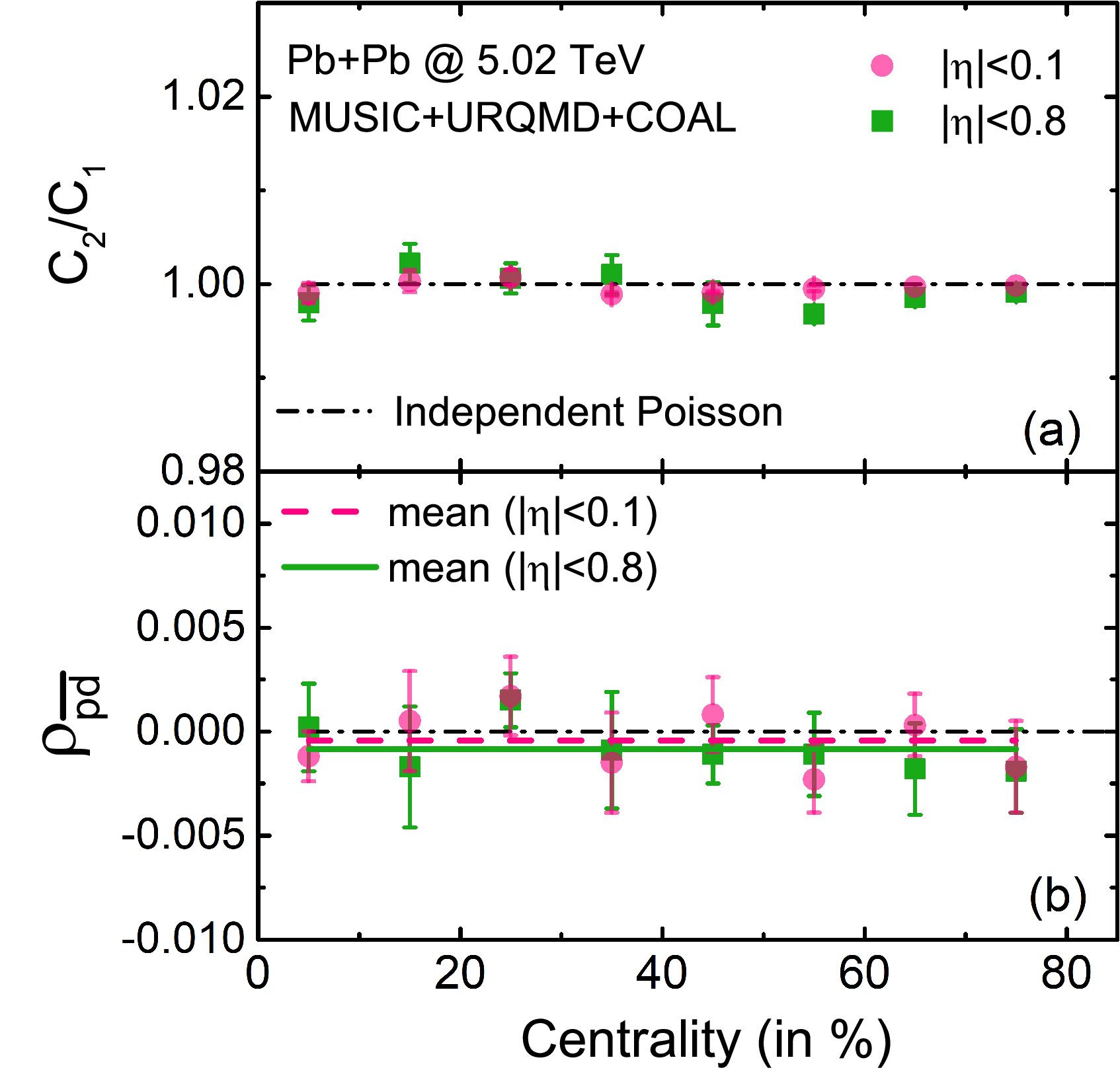}
  \caption{Centrality dependence of the scaled moment $C_2/C_1$ (panel (a)) as well as the correlation between antiproton and anti-deuteron multiplicities (panel (b)) in Pb+Pb collisions at $\sqrt{s_{NN}}=5.02$~TeV for different pseudorapidity acceptance windows.
  }
  \label{pic:cen}
\end{figure}

We further display in Fig.~\ref{pic:cen} the centrality dependence of the scaled moment $C_2/C_1$ (panel (a)) and also the correlation between the antiproton and anti-deuteron  multiplicity distributions (panel (b)) in Pb+Pb collisions at $\sqrt{s_{NN}}=5.02$~TeV for different pseudorapidity acceptance windows. The values of $C_2/C_1$ are consistent with the Poisson limit  for all values of the acceptance window.   The correlation between anti-deuteron and antiproton multiplicity distributions is consistent with zero within uncertainties. By taking an average of $\rho_{\bar{p}\bar{d}}$ over all centralities, we have found a small negative mean value of around $-4.2\times 10^{-4}$ and $-8.5\times 10^{-4}$ for $|\eta|<0.1$ and $|\eta|<0.8$, respectively. Baryon conservation tends to yield a negative $\rho_{\bar{p}\bar{d}}$, while a positive correlation between antiproton and antineutron would give a positive  $\rho_{\bar{p}\bar{d}}$. The small negative $\rho_{\bar{p}\bar{d}}$ is a result of these two competing effects.

\emph{Summary.}{\bf ---} Event-by-event multiplicity fluctuations and correlations have been suggested  as a sensitive probe to the production mechanism of fragile anti-nuclei in high-energy nuclear collisions. In the present study, we have investigated the event-averaged yield of anti-deuterons and its event-by-event fluctuation using the nucleon coalescence model for their production from anti-nucleons at the kinetic freeze-out of a microscopic hybrid  approach based on the MUSIC hydrodynamic model and the UrQMD hadronic transport model.   We have found  a suppression of (anti-)deuteron production in peripheral collisions of Pb+Pb collisions at $\sqrt{s_{NN}}=5.02$~TeV, which is in accordance with the experimental measurements from the ALICE Collaboration. We have also found that the scaled moment $C_2/C_1$ of the anti-deuteron multiplicity distribution obtained from our coalescence model calculation agrees with the Poisson limit for a grand canonical ensemble but is smaller than that obtained from the simple coalescence model that assumes  the same probability for all antiproton and antineutron pairs to form  deuterons, which  leads to an excess over the Poisson limit by about $2\langle N_{\bar{d}}\rangle/\langle N_{\bar{p}}\rangle$.  Moreover, we have found a small negative correlation between the anti-deuteron and  antiproton multiplicity distributions with a mean value of around $-8.5\times 10^{-4}$  for the pseudorapidity acceptance window $|\eta|<0.8$ after averaging over all centralities. These results provide quantitative references for making comparisons with experimental measurements and for understanding the production mechanism of light (anti-)nuclei  in high-energy nuclear collisions.

\begin{acknowledgments}
We thank Maximiliano Puccio, Alexander Philipp Kalweit, Sourav Kundu, Benjamin D\"onigus, Xiaofeng Luo, Rui Wang, and Wenbin Zhao for helpful discussions. This work was supported in part by the U.S. Department of Energy under Award No.DE-SC0015266.
\end{acknowledgments}

\medskip
\appendix~~~~~~~~~~~~~~~~~~~~~~~~~~~~Appendix
\medskip

\emph{Scaled moment of anti-deuteron multiplicity distribution in a simple coalescence model.}{\bf ---} In Model B of the simple coalescence model  in Ref.~\cite{Feckova:2016kjx}, the antiproton and antineutron multiplicities in each event are assumed to follow independent Poisson distributions, i.e., $N_{\bar{p}}~\sim~Pois(\langle N_{\bar p}\rangle)$ and $N_{\bar{n}}~\sim~Pois(\langle N_{\bar n}\rangle)$. The anti-deuteron multiplicity $N_{\bar{d}}$ then follows a Binomial distribution, i.e., $N_{\bar{d}}\sim B(N_{\bar{n}}N_{\bar{p}}, \langle N_{\bar d}\rangle^\prime/(N_{\bar{n}} N_{\bar{p}}))$, where   $\langle N_{\bar d}\rangle^\prime$ is the mean number  of anti-deuteron given $N_{\bar{n}}$ neutrons and $N_{\bar{p}}$ protons. Denoting the event-averaged value of $\langle N_{\bar d}\rangle^\prime$ as $ \langle N_{\bar d}\rangle$ and using the properties of Poisson distribution and Binomial distribution, one can obtain 
\begin{eqnarray}
\frac{C_2}{C_1}\approx 1+\frac{\langle N_{\bar{d}}\rangle}{\langle N_{\bar{n}}\rangle}+\frac{\langle N_{\bar{d}}\rangle}{\langle N_{\bar{p}}\rangle}\nonumber  
\end{eqnarray}
For central Pb+Pb collisions at $\sqrt{s_{NN}}=5.02~$TeV, we find $\langle N_{\bar{d}}\rangle/\langle N_{\bar{n}}\rangle\approx \langle N_{\bar{d}}\rangle/\langle N_{\bar{p}}\rangle$ and thus $C_2/C_1\approx 1+2\langle N_{\bar{d}}\rangle/\langle N_{\bar{p}}\rangle$.


\end{document}